
\magnification=1200 \def\wc{\hangindent=4em \hangafter=1 \noindent}
\baselineskip 14pt \parskip 3pt \null 
\headline={\ifnum\pageno=1\hfil\else\hfil\tenrm--\ \folio\ --\hfil\fi}
\footline={\hfil}

\centerline{\bf THE COSMIC MICROWAVE BACKGROUND DIPOLE AS A COSMOLOGICAL
EFFECT}
\vskip 1.0cm
\centerline{M. Jaroszy{\'n}ski$^{1,2}$ and B. Paczy\'nski$^2$ }
\vskip 0.5cm

\centerline{ $^1$ Warsaw University Observatory, Al. Ujazdowskie 4,
                00-478 Warsaw, Poland}
\centerline{ $^2$ Princeton University Observatory, Peyton Hall,
                Princeton, NJ 08544 USA}
\vskip 1.0cm
\centerline{\it Received: ...........................................}
\vskip 0.5cm
\centerline {\bf ABSTRACT}
A conventional explanation of the dipole anisotropy of the cosmic
microwave background (CMB) radiation is in terms of the Doppler effect:
our galaxy is moving with respect to CMB frame with $ \sim 600 ~ km ~ s^{-1} $.
However, as the deep redshift surveys fail to reveal a convergence of
the large scale flow to zero at distances as large as $ d \sim H^{-1}
15,000 ~ km ~ s^{-1} $ (Lauer \& Postman, 1994),
the uniqueness of the conventional interpretation
has to be investigated.  A possible alternative might be a cosmological
entropy gradient, as suggested by Paczy\'nski \& Piran (1990).  We find
that contrary to that suggestion a quadrupole anisotropy is generically
of the same order of magnitude as the dipole anisotropy (or larger) not
only for adiabatic but also for iso-curvature initial perturbations.
Hence, the observed dipole cannot be explained with a very large scale
perturbation which was initially iso-curvature.
\vskip 0.5cm

\vskip 0.5cm
\wc{{\it Subject headings:} cosmic background radiation - cosmology -
gravitation}

\vfill\eject

\centerline {\bf 1. INTRODUCTION}
\vskip 0.5cm

The dipole moment of the CMB is usually interpreted to be the result of a
Doppler effect caused by our motion with respect to CMB frame (cf. Kogut
{\it et al.} 1993
and references therein).  According to this interpretation the CMB,
as well as galaxies (when averaged over a large enough volume) define the
local standard of rest.  Our galaxy, together with its neighbours moves
with respect to the local standard of rest.  When the velocities of galaxies
in a large enough volume are measured they should be found to be at
rest, naturally after the allowance is made for the overall Hubble expansion.

The observations available so far fail to provide a clear support for this
picture.  The recent most troublesome result is that of Lauer \& Postman
(1994) who find that the frame defined with the 119 Abell clusters of galaxies
within $ 15,000 ~ km ~ s^{-1} $ is moving at $ \sim 700 ~ km ~ s^{-1} $
with respect to the local standard of rest as defined by the CMB.  This
scale is so large that it is difficult to accomodate within most currently
available models of the formation of large scale structure in the universe
(Strauss et al. 1994).  This trouble persisted for many years (cf. Paczy\'nski
\& Piran 1990, hereafter PP, and references therein) and it justifies
a search for alternative interpretations of the CMB dipole anisotropy.
PP proposed that entropy gradient on a scale larger than the current horizon
could give rise to a dipole moment while keeping the quadrupole unmeasureably
small.

The purpose of this paper is to demonstrate that the PP proposal was incorrect.
PP used Tolman-Bondi cosmological model with no pressure to demonstrate
that the very long wavelength density perturbations give rise mostly to
a quadrupole anisotropy, while entropy perturbations show up as a dipole.
The reason for ignoring pressure was the currently very small value of
density and pressure due to CMB.  The CMB was dynamically important
only at redshifts larger than $ z_{eq} \sim 10^4 $, and it seemed safe to
ignore it, or at least it seemed to be a reasonable first approximation.
It turns out that was a conceptual mistake.  The purpose of this paper
is to demonstrate that no matter how small is the current contribution of
radiation to the closure density any large scale iso-curvature (entropy)
perturbation generically gives rise to a quadrupole CMB anisotropy
which is larger than the dipole anisotropy.  We demonstrate this in
section 2 for a plane wave, and in section 3 for a modified Tolman-Bondi
model.  Finally, a simple qualitative demonstration of the fact that
the initially iso-curvature (entropy) perturbation generically gives rise
to a density perturbation (cf. Peebles, 1993, hereafter P93)
is presented in section 4, together with a discussion
of other possibilities.

\vfill \eject

\centerline {\bf 2. PLANE WAVE PERTURBATIONS IN A FLAT UNIVERSE}
\vskip 0.5cm

\noindent {In the synchronous gauge (P93)
the perturbed metric
of the flat universe model has the form:}
$$
ds^2=c^2dt^2-a^2(t)(\delta_{jm}-h_{jm})dx^jdx^m
\eqno (2.1)
$$
where $t$ is the cosmic time, $a(t)$ the scale factor in the unperturbed
model, $\delta_{jm}$ is the Kronecker delta symbol and $h_{jm}(t,x^l)$
are the small perturbations to the metric. The spatial coordinates are
Cartesian, Latin indices go through 1,2,3.
We are interested in the growing, scalar modes of perturbations to the metric
(Lifschitz, 1946). In the case of the plane wave perturbation
with the wave vector $\bf k$ along the $x^3$ axis, the only nonvanishing
components of $h_{jm}$ are the diagonal terms and $h_{11}=h_{22}$
because of the symmetry. Thus there are two independent variables
defining tensor $h_{jm}$ and we choose $h=h_{11}+h_{22}+h_{33}$
and $H=h_{11}-h_{33}$ for the purpose.

In our model the matter consists of two independent components, which interact
which each other by gravity only. The first component is non relativistic
(NR) matter (for example baryons) and it is characterised by the present
dimensionless density $\Omega_{NR}=\rho_{NR}/\rho_c$,
where $\rho_c \equiv 3H_0^2/8\pi G$ is the critical cosmological density,
$H_0$ is the Hubble's constant and $\rm G$ is the constant of gravity.
The second component
is ultra-relativistic (UR) matter (for example electromagnetic radiation)
and it is characterised by the present dimensionless
density $\Omega_{UR}=\rho_{UR}/\rho_c$.
Since we are using the flat unperturbed cosmological model, we add
the cosmological constant to be in agreement with unperturbed evolutionary
equations:
$$
\left({{\dot a} \over a} \right)^2 =
{8\pi G \over 3c^2}(\epsilon_{NR}+\epsilon_{UR})
+{1 \over 3} \Lambda c^2
\eqno (2.2)
$$
$$
{{\ddot a} \over a} =
-{4\pi G \over 3c^2}(\epsilon_{NR}+\epsilon_{UR}+3P_{UR})
+{1 \over 3} \Lambda c^2
\eqno (2.3)
$$
where dots mean the time differentiation,
$\epsilon_i \equiv \rho_i c^2$ depict the full energy densities of various
components and $P_i$ - their pressures.
We introduce dimensionless cosmological constant
$\lambda={1 \over 3}\Lambda c^2/H_0^2$. In a flat model we have:
$$
\Omega_{NR}+\Omega_{UR} + \lambda = 1
\eqno (2.4)
$$
In some cases $\lambda$ may be negative.

Throughout this paper we assume that all perturbations of
all non-relativistic matter follow strictly the perturbations of baryons.
We also assume that all perturbations of all ultra-relativistic matter
follow strictly the perturbations of electromagnetic radiation.
The ultra-relativistic and non-relativistic components are coupled
to each other prior to recombination, and they become decoupled
when the universe becomes transparent to radiation at $ z_{rec} \gg 1 $.

The energy density fluctuations of different components are described by:
$$
{\delta \rho_{NR} \over \rho_{NR}} = \delta_{NR} ~~~~
{\delta \rho_{UR} \over \rho_{UR} }
= {4 \over 3}\delta_{UR}
 \eqno(2.5)
$$
where $\delta_{NR}$, $\delta_{UR}$ are the relative perturbations in particle
density of the two componets.
The perturbations in specific entropy
$\delta_S \equiv \delta S/S$, where $S$ is the entropy of
ultrarelativistic fluid per particle of nonrelativistic fluid
can be exppressed as $\delta_{S}=\delta_{UR}-\delta_{NR}$.
(We neglect the entropy of the nonrelativistic fluid).
The velocity perturbation of each fluid has the form
$u^\alpha = (1/c,v^j/ac)$ where $\bf v$ is the physical
velocity measured by the synchronous observers . In the scalar mode
of perturbations only the correction to the expansion, $\Theta$,
enters the equations:
$$
u^\alpha_{;\alpha} = 3{{\dot a} \over a} + \Theta~~~~
\Theta = {{\dot a} \over a} \chi
\eqno(2.6)
$$
The dimensionless variable
$\chi$ is more convenient than $\Theta$ (Press \& Vishniac, 1980,
hereafter PV80)
and we need $\chi_{NR}$ and $\chi_{UR}$ to characterize both components.

Our set of equations describing the evolution of small perturbations
to the metric and fluid variables is based on P93 and PV80.
We use a new time coordinate (PV80)
$\eta \equiv ln(a/a_0) \equiv -ln(1+z)$,
where $a_0$ is the present characteristic scale in the Universe
and $z$ is the redshift.
The spatial gradients are already omitted in the equations, so they
are valid for very large scale perturbations only. The equations
are written for two fluid interacting only gravitationally (but see
below):
$$
h^{\prime\prime} + \left({{\ddot a}a \over {\dot a}^2}+1 \right)h^{\prime}
={8\pi G a^2 \over {\dot a}^2}
\left(\rho_{NR} \delta_{NR}+{8 \over 3}\rho_{UR}\delta_{UR}
\right)
\eqno(2.7)
$$
$$
H^{\prime\prime} + \left({{\ddot a}a \over {\dot a}^2}+2 \right)H^{\prime}
=-{8\pi G a^2 \over {\dot a}^2}
\left(\rho_{NR} \delta_{NR}+{4 \over 3}\rho_{UR}\delta_{UR}
\right) -h^{\prime}
\eqno(2.8)
$$
$$
\delta_{NR}^{\prime}=({1 \over 2}h^{\prime}- \chi_{NR})
\eqno(2.9)
$$
$$
\delta_{UR}^{\prime}=({1 \over 2}h^{\prime}- \chi_{UR})
\eqno(2.10)
$$
$$
\chi_{NR}^\prime+
\left({{\ddot a}a \over {\dot a}^2} +1 \right)\chi_{NR}=0
\eqno(2.11)
$$
$$
\chi_{UR}^\prime+\left({{\ddot a}a \over {\dot a}^2}
 \right)\chi_{UR}=0
\eqno(2.12)
$$
where we have neglected sound velocity  in the nonrelativistic component
and for the relativistic component
we have already substituted $1/3$ for its pressure
to energy density ratio and for its sound velocity square.
The time derivatives of $a$ can be substituted from Eqs. (2.2), (2.3).

We solve our equations for different values of the universe model parameters
varying the density of non-relativistic
component in the limits $0.01 \le \Omega_{NR} \le 1$
and the density of ultra-relativistic component is in the range
$10^{-5} \le \Omega_{UR} \le 1$.
The value of cosmological constant results from Eq. (2.4).

We always set initial conditions long before the time of recombination
and long before the time when the energy density of the ultra-relativistic
component
is equal to the energy density of the non-relativistic component,
whichever comes first. In this early time the radiation dominated plasma
behaves like a relativistic fluid and the unperturbed model evolves
like a model with relativistic equation of state.
(Various terms in the equations have different redshift dependence.
While $ \rho_{UR} \sim (1+z)^4$ and $\rho_{NR} \sim (1+z)^3$ the terms
including the cosmological constant do not depend on the redshift. Thus
the effects of $\Lambda \ne 0$ can only show when $z$ is small.)
In the early times the coefficients in the above equation set remain constant
so one may find independent modes of perturbations (Lifshitz, 1946, P93 and
references therein). We consider two types of initial conditions.
The first, {\it adiabatic} perturbation, has three possible modes in the
relativistic Universe.
The most natural of them , with growing rate proportional to
$a^2$, remains regular at $t \rightarrow 0$ and has vanishing velocity
($\chi_{NR} \equiv 0$, $\chi_{UR} \equiv 0$).
Another growing  mode in the early epoch
with amplitudes $\sim a^1$ has nonvanishing velocity
and irregular some of the metric components when $t \rightarrow 0$ (P93).
Since the rate of instability in this mode is slower as compared to the
previous one, it can not appreciably influence the present Universe
unless the initial conditions are fine tuned.
The decaying mode is of no interest: first it cannot produce any significant
perturbation to the present Universe, second it is unphysical (PV80,
Bardeen, 1980).
For {\it adiabatic} perturbations we have $\delta_{NR}=\delta_{UR}$
initially and this condition is preserved if velocities vanish.
(This mimics the coupling between ordinary matter and radiation, if
required, without explicitly puting it into equations).

The second, {\it isocurvature} type of perturbations (Peebles, 1987)
is impossible in a single fluid model.
With more than one fluid we are able to introduce a spatial dependence
of the chemical composition of the matter not introducing perturbations to the
energy density or to the metric. We just perturb the density of
non-relativistic component, not
changing the density of ultra-relativistic component. As long as
the relativistic component dominates, no energy density perturbation arises
and perturbations change slowly ($\delta_{NR} \approx const$,
$h^\prime$, $H^\prime$, $\delta_{UR} << \delta_{NR}$). When nonrelativistic
component becomes dynamically important the growing energy density
perturbation results. Examining  Eqs. (2.9) and (2.10) we see, that the
entropy perturbation $\delta_{S}=\delta_{UR}-\delta_{NR}$ remains
constant.

The two kinds of perturbations considered behave differently
at the begining, but become similar when the model becomes nonrelativistic.
We are interested in the fluctuations in the microwave background caused
by the perturbations. They arise between the surface of last scattering
and the present epoch. To compare results in different models we normalize
all gravitational instability calculations in such a way that
the present amplitude of the density perturbations of non-relativistic
component is the same and has the value $\delta$.
According to the above remarks it is not surpraising that the influence of
both kinds of perturbations on the microwave background is
similar.

To find the temperature of the CMB radiation in any particular direction
on the sky $\bf n$, we have to follow rays back in this direction.
The coordinate distance travelled by a particle moving with the
velocity of light is given by
$$
\tau(t)=\tau(t(\eta))=\int^t{c~dt \over a}
=\int^\eta{c~d\eta \over {\dot a}}
\eqno(2.13)
$$
The position of a photon, which is now at ${\bf x}_0$,
coming from the direction $\bf n$, was  at the ``time'' $\eta$:
$$
{\bf x}_\eta={\bf x}_0+{\bf n}(\tau_0-\tau_\eta)
\eqno(2.14)
$$
where we have substituted $\eta=0$ for the present epoch. $\tau_0$ measures
the present coordinate distance to the horizon.

The temperature fluctuation of the CMB radiation coming to the observer
at ${\bf x}_0$ from the direction $\bf n$
on the sky is given as (Sachs and Wolfe, 1967):
$$
\left({\delta T \over T}\right) ({\bf x}_0,{\bf n}) =
\left({\delta T \over T}\right)_{\rm rec}+{1 \over c}
\left({\bf v}_{\rm obs}-{\bf v}_{\rm rec}\right){\bf n}
- {1 \over 2} \int_{\eta_{rec}}^0
{h_{jm}^\prime(\eta,{\bf x}_\eta) n^j n^m~d\eta}
\eqno(2.15)
$$
where the subscript ``${\rm rec}$''
 denotes the quantities measured at the epoch of recombination, at the
place from which rays come.
The first term coresponds to the temperature fluctuations in the plasma
at the last scattering surface. It is equal to ${1 \over 3}\delta_{UR}$
in the place of emission.
For a plane wave with a wave vector $\bf k$
we define the directional cosine $\mu=\cos \theta ={\bf n}{\bf k}/k$.
In our case $\bf k$ is along the $x^3$ axis and phase factor along the
ray changes like
$exp(ikx^3+ik \mu (\tau_0-\tau))$.
$k(\tau_0-\tau_{rec}) \approx k\tau_0 << 1$ for superhorizon perturbations.
Expanding and taking real part we have to the lowest order:
$$
\left({\delta T \over T}\right)_{\rm rec}
=-{1 \over 3}~k\tau_0~\sin(kx^3)~\delta_{UR}(\eta_{rec})~P_1(\mu)
\eqno (2.16)
$$
$P_1(\mu)=\mu$ is the Legendre polynomial of the first order and the above
expression contributes to the dipole anisotropy of the CMB (this
contribution we shall denote $D_{\rm rec}$).
 The quadrupole part ($Q_{\rm rec}$)
is of still higher order being proportional to $(k\tau_0)^2$.

The second term is due to the Doppler shift caused by the difference
in velocity of matter between recombination and the present epoch.
Since we do not consider velocity perturbations, ${\bf v}_{\rm rec}$
vanishes automatically. The observer peculiar velocity ${\bf v}_{\rm obs}$
may arise from the small scale perturbations to the gravitational field
but is of no interest here.

The third term is the result of gravitational field acting on photons
(the Sachs - Wolfe effect).  With our definitions of $h$, $H$ we have:
$$
{1 \over 2}~h_{jm}^\prime~n^j n^m =
{1 \over 6}~{h^\prime}~P_0(\mu) - {1 \over 3}~{H^\prime}~P_2(\mu)
\eqno(2.17)
$$
where $P_n(\mu)$ are the Legendre polynomials .\footnote*{
In general case also the amplitudes of {\it vector} metric perturbations
coupled to the spherical harmonics $Y^2_{\pm 1}$ and the {\it tensor}
amplitudes times $Y^2_{\pm 2}$ would appear in eq.(2.17)}

The following of rays is required in the integration of  Eq.(2.16).
 The gravitationally induced
temperature fluctuations are given as:
$$
\left({\delta T \over T}\right)_{\rm SW} (\mu)
=-{1 \over 6}e^{ikx^3}~
 \int_{\eta_{rec}}^0~
\left(h^\prime(\eta)~P_0(\mu)-2H^\prime(\eta)~P_2(\mu)\right)~
 ~e^{ik \mu (\tau_0-\tau_\eta)}~d\eta
\eqno(2.18)
$$
For a superhorizon perturbation the exponent
under the integral can be expanded. We get the series in the small
quantity $k\tau_0$ with coefficients being the
products of $P_0$ and $P_2$ with powers of $\mu$.
We limit ourselves to the dipole and quadrupole terms:
$$
D_{\rm SW}=-k\tau_0~\sin(kx^3)~\int_{\eta_{rec}}^0
{}~\left({1 \over 6}h^\prime(\eta)-{2 \over 15}H^\prime(\eta) \right)~
\left(1-{\tau \over \tau_0} \right)~d\eta
\eqno (2.19)
$$
$$
Q_{\rm SW}=
-\cos(kx^3)~\int_{\eta_{rec}}^0~{1 \over 3}H^\prime(\eta)~d\eta
\eqno (2.20)
$$

One can define the density difference accross the horizon
$\Delta=|\tau_0\nabla \delta| \approx k\tau_0 \delta$. For various
contributions
to anisotropy we have:
$$
D_{\rm rec} = d_{\rm rec}~\Delta ~~~~~
D_{\rm SW}  = d_{\rm SW} ~\Delta
\eqno(2.21)
$$
and
$$
Q_{\rm rec} = q_{\rm rec}~{\Delta^2 \over \delta}~~~~~
Q_{\rm SW}  = q_{\rm SW}~\delta
\eqno(2.22)
$$
In Fig.1 we show the ratio of the dipole to quadrupole CMB anisotropy
($D/Q \approx D_{\rm SW}/Q_{\rm SW}$ measuring it by the ratio of the
density difference through the horizon to the density perturbation amplitude
($\Delta/\delta$) which is a small quantity being the ratio of the present
horizon size to the present scale of perturbation. The dipole anisotropy
caused by the superhorizon perturbation is always smaller than the resulting
quadrupole as can be seen on the plots.

The expansion of the perturbed model is not isotropic and the ``Hubble
constant''  depends on the direction of measurement.
Comparing the proper distance to a close object in the direction $\bf n$
with its
velocity due to the expansion one gets:
$$
H({\bf n})={{\dot a} \over a} - {1 \over 2}{\dot h}_{lm}n^ln^m
={{\dot a} \over a} \left[1- {1 \over 6}h^\prime(0,x^3)P_0(\mu)
      +{1 \over 3}H^\prime(0,x^3)~P_2(\mu) \right]
\eqno(2.23)
$$
where the variables are calculated at the present time.
The monopole part is of no interest. Perturbations in the metric introduce
the quadrupole anisotropy to the Hubble law with the relative amplitude:
$$
Q_{\rm H} = {1 \over 3}cos(kx^3)~H^\prime(0)
\eqno(2.24)
$$
As one can see, the Hubble anisotropy has the same phase of spatial dependence
as the quadrupole anisotropy of CMB, but the amplitudes are defined
by different quantities.

\vskip 1.0cm
\centerline{\bf 3. SPHERICAL SOLUTIONS}
\vskip 0.5cm

We follow here the approach to the spherically symmetric world models
outsketched in the Appendix B of PP.
As in the previous chapter and generally, when the inhomogeneities have
the scale much larger than the horizon scale at the epoch of interest,
we are going to neglect the influence of pressure gradients on the
evolution of the model. We use the Bondi-Tolman metric:
$$
ds^2=c^2dt^2-X^2(t,r)dr^2-R^2(t,r)d\Omega^2
\eqno (3.1)
$$
where r is a radial coordinate and $d\Omega^2$ is the line
element on the sphere.
(In fact the presence of pressure gradients forces one to use a more general
spherically symmetric metric with $g_{tt} = A^2(t,r)$, since
the field equations say that the pressure gradient is proportional
to the gradient of $A(t,r)$, as shown by May \& White (1967).
Thus, strictly speaking, the synchronous,
comoving coordinate system is impossible
in the presence of pressure gradients.)
Neglecting pressure gradients from the begining, we adopt line element (3.1)
in our calculations.

In the case of a single fluid one can choose the coordinate system (3.1) to
be comoving with the matter, so the velocities vanish automatically. In the
case of many fluids interacting only gravitationally it is possible that
pressure gradients, not necessarilly equal in different components, may
cause the relative motion of the fluids. But we are neglecting pressure
gradients from the begining, so it is fair to assume that relative motions
of the fluids are impossible. Thus it is possible to make
the coordinate system comoving with the matter.
In that case the mixed components of the energy-momentum tensor vanish
automatically ($T^{tr} \equiv 0$) and, as a consequence of the field
equations one has:
$$
X={R_{,r} \over W(r)}
\eqno (3.2)
$$
where $W(r)$ is a free function (Bondi, 1947, PP). Equation (3.2) implies,
that one can write down the field equation for $R(t,r)$ in the form:
$$
{\dot R}^2=W^2(r)-1+{2Gm(t,r) \over R}
\eqno (3.3)
$$
where
$$
m(t,r)= 4\pi \int_0^r {\rho R^2R_{,r}dr}
=4\pi \int_0^r{\rho R^2XWdr}
\eqno(3.4)
$$
The quantity defined above is the {\it gravitational} mass in the sphere
inside $r$. The second integral shows that it is the total matter density
integrated over the proper volume with some ``weighting'' function
$W(r)$. Equations with the form of
Eqs (3.3)and (3.4) are valid in the general case of spherically
symmetric configuration with pressure (May \& White, 1967).
Equation (3.3) has the form of energy equation for a test particle
in the field of [possibly variable] mass $m$.
While the density distribution inside the configuration defines
the time dependent potential, the function $W(r)$ sets the initial
velocity of the particle.

If the pressure vanishes, mass inside any radius $r$ is preserved
and Eq (3.3) becomes an ordinary differential equation with
coordinate $r$ playing the role of a parameter (PP).
In the general case of matter having an admixture of relativistic
component (at least photons are such a component) one may solve the
evolutionary equation for $R(t,r)$ using the following approach.
First we divide the configuration into a number of concentric, thin
shells. The innermost spherical region behaves like a part of uniform
solution. For this region one has:
$$
m_{NR}^1(t)=m_{NR}^1(t_{init})
{}~~~~m_{UR}^1(t)=m_{UR}^1(t_{init}){R(t_{init},r_1) \over R(t,r_1)}
\eqno(3.5)
$$
where $m_{NR}^1$, $m_{UR}^1$ are the masses of all non-relativistic and
ultra-relativistic components,
respectively, inside the first zone. The same behavior is true for any other
non-relativistic/ultra-relativistic components.
During adiabatic expansion a relativistic fluid changes energy density as
$\rho_{UR} \sim V^{-4/3}$ where $V$ is the proper volume of the region.
Using Eq. (3.4) we have for a thin shell of matter between $r_j$
and $r_{j+1}$:
$$
m_{UR}^{j+1}(t)-m_{UR}^{j}(t)=
\left[m_{UR}^{j+1}(t_{init})-m_{UR}^{j}(t_{init})\right]
\left[{R^3(t_{init},r_{j+1})-R^3(t_{init},r_{j}) \over
       R^3(t,r_{j+1})-R^3(t,r_{j})}\right]^{1 \over 3}
\eqno (3.6)
$$
where we implicitly assumed that $W(r)$ can be treated as constant
through a single zone.
Using Eq. (3.5) one can obtain solution for $R(t,r_1)$ and $m(t,r_1)$.
Knowing $R(t,r_j)$ and $m(t,r_j)$ one can find $m(t,r_{j+1})$ as a function
of $R(t,r_{j+1})$ with the help of Eq. (3.6). Substituting this dependence
sinto Eq. (3.5) one finds solution for $R(t,r_{j+1})$. Recurently one
can find the metric functions values and their derivatives on a grid.
In other points metric can be found by interpolation.

We start calculations early, when ultra-relativistic component of matter
dominates the dynamics ($\rho_{UR} >> \rho_{NR}$.
At this stage we always set relativistic fluid density to be constant
in space. The perturbation is in the free function $W(r)$, which we borrow
from PP:
$$
W^2(r)=1-{r_0^2-r^2 \over r_0^2+r^2}~{r^2 \over r_0^2}
\eqno(3.7)
$$
where $r_0>>1$ sets the spatial size of perturbation. In our convention
coordinate distance $r \approx 1$ corresponds to the present horizon size.
The density of non-relativistic component is also set to constant in the case
of adiabatic perturbations. For entropy perturbations we use the
following shape of initial density of the non-relativistic component:
$$
\rho_{NR} \sim \Omega_{NR}~\left[2-\left({r \over 2r_0}\right)^3\right]
{}~~~~r \le 2r_0
\eqno(3.8)
$$
Accordingly the specific entropy behaves like:
$$
S(r) \sim \rho_{NR}^{-1} \sim {1 \over 2-\left({r \over 2r_0}\right)^3}
\eqno(3.9)
$$

We put observers at different places in the world models described above.
The comoving volume is defined as:
$$
V(t,r)={R^2(t,r)X(t,r) \over r^2}
\eqno (3.10)
$$
Observer at any location $(t_{\rm obs},r)$, where $t_{\rm obs}$ is
to represent the epoch of observation,
can define the recombination epoch time $t_{\rm rec}$ using
the condition:
$$
V(t_{\rm obs},r)=(1+z_{\rm rec})^3V(t_{\rm rec},r)
\eqno(3.11)
$$
where $z_{\rm rec} \approx 10^3$ is the redshift factor of recombination.
Suppose we follow a ray, which goes along a null geodesics
from the last scaterring surface
at $(t_{\rm rec},r_{\rm rec})$ to the observer
at $(t_{\rm obs},r_{\rm obs})$.
The starting point of the photon depends on the direction $\bf n$ from which it
arrives, $r_{\rm rec}=r_{\rm rec}({\bf n})$.
The locally measured energies of a photon
would be $E_{\rm rec}$ and $E_{\rm obs}$ respectively, the comoving volumes
at the locations - $V_{\rm rec}$ and $V_{\rm obs}$. Since initially
the radiation temperature was constant through the space and it changes
accordingly to the law $T \sim V^{-1/3}$, we can define it at any time
and location. Taking into account the redshift of photons between
last scattering surface and the observer we have:
$$
T({\bf n})=T_{\rm local}~{E_{\rm obs} \over E_{\rm rec}}~
\left({V_{\rm obs} \over V_{\rm rec}}\right)^{1 \over 3}
\eqno (3.12)
$$
where $T_{\rm local}$ is the locally measured, average temperature of the CMB.
We compare results of such CMB temperature measurement  in the direction
radially in ($T_1$), radially out ($T_3$) and in transverse direction ($T_2$).
The average temperature on the sky in our approximation is given as:
$$
T_{\rm local}={1 \over 4}(T_1+2T_2+T_3)
\eqno (3.13)
$$
We define the dipole  and quadrupole anisotropy of the CMB as:
$$
D={T_1-T_3 \over 2~T_{\rm local}}~~~~
Q={T_1-2T_2+T_3 \over 4~T_{\rm local}}
\eqno (3.14)
$$
According to our calculations
the dipole is always about 2 orders of magnitude weaker
as compared to the quadrupole. The exception are the places, where
the quadrupole vanishes locally, but such places are rare and the
a'priori probability of making observations from there is low.

\vskip 1.0cm
\centerline{\bf 4. DISCUSSION}
\vskip 1.0cm

In sections 2 and 3 we presented a formal demonstration that a perturbation
which is initially iso-curvature (i.e. the entropy is perturbed)
while the ultra-relativistic component dominates, grows into a curvature
(density) perturbation when the overall expansion of the universe makes
the non-relativistic component dominant.  The dynamics of the universe is
believed to be currently dominated by the non-relativistic component
(e.g. baryonic, or cold dark matter), while it was dominated in the past by
the ultra-relativistic component (e.g. radiation, pairs, etc.).  We can
envision the expansion history of two universes which differ in the initial
entropy per baryon, or any other measure of the ratio of ultra-relativistic
to non-relativistic components, both universes being exactly flat, isotropic
and homogeneous.
\footnote*{Similar reasoning can be found in Tolman, 1934 and P93.}

The two expansion histories are synchronized at $ t_0 = 0 $.  Initially,
they are identical, with the scale factor increasing
as $ t^{1/2} $ while the dynamics is dominated by the ultra-relativistic
component.  The expansion rate changes to $ a \sim t^{2/3} $ at the time
$ t = t_{eq} $, when the non-relativistic component becomes dominant.  This
time is different for the two universes with the different ratio of the
two components, and therefore, the late expansion rate is different too.
This means that the scale factors for the two universes are not the same
at any time $ t > t_{eq} $.  If these two universes are just large parts of
the same universe and they come into contact at $ t > t_{eq} $, the
mis-match of their scale factors will create space curvature, which
will give rise to a quadrupole anisotropy of the CMB.
On Fig. 2 we show the expansion histories of different parts of our
spherical model. At the begining $R(t,r)/r  \sim t^{1/2}$ independent
of position. In later times the differences become visible.

Our conclusion is that the dipole anisotropy of the CMB cannot be explained
with a very large scale entropy gradients in the universe, as proposed by
Paczy\`nski \& Piran (1990).  If the dipole is cosmological in origin
something else is needed.  We see no `natural' explanation.  A trivial and
entirely ad hoc and artificial `explanation' can be offered.  Imagine
there are many ultra-relativistic components, with one having a very
large scale fluctuation which is exactly in the opposite phase than
radiation.  There would be no overall change in the ratio of ultra-relativistic
to non-relativistic componet, and no effect on the dynamics of the universe,
as any effect of the radiation perturbation would be exactly balanced by
the opposite effect due to another ultra-relativistic component, yet there
would be a dipole in the CMB.  There is no justification for such a proposal,
and we mention it as an example of what may have to be considered if the
Lauer \& Postman (1994) result is confirmed and even extended to ever larger
scales.

This project was supported with the NSF grant AST 93-13620, NASA grant
NAG5-2759 and KBN grant 2~P304~006~06.

\vskip 1.0cm
\centerline {\bf REFERENCES}
\vskip 0.5cm

\wc{Bardeen, J.M. 1980 Phys.Rev.D, 22, 1882 \hfill}

\wc{Bondi, H., 1947, MNRAS, 107, 410 \hfill}

\wc{Kogut, A. {\it et al.}, 1993, Apj, 419, 1. \hfill}

\wc{Lauer, T., \& Postman, M.  1994, ApJ, 364, 341 \hfill}

\wc{Lifschitz, E.M. 1946 , J.Phys (Moscow), 10, 116. \hfill}

\wc{May  \& White  1967, Meth.Comp.Phys., 7, 219 \hfill}

\wc{Paczy\'nski, B., \& Piran, T.  1990, ApJ, 364, 341 (PP)  \hfill}

\wc{Peebles, P.J.E. 1987, ApJ, 315, L73 \hfill}

\wc{Peebles, P.J.E. 1993, in: {\it Principles of Physical Cosmology},
 Princeton University Press, Princeton.(P93) \hfill}

\wc{Press, W.H. and Vischniac, E.T. 1980, ApJ, 239, 1 (PV80) \hfill}

\wc{Sachs, R.K. and Wolfe, A.M. 1967, ApJ, 147, 73. \hfill}

\wc{Strauss, M. A., Cen, R., Ostriker, J. P., \& Lauer, T. R. 1994,
submitted to ApJ, Princeton Observatory Preprint 572  \hfill}

\wc{Tolman, R.C. 1934, Proc.Nat.Acad.Sci., 20, 169 \hfill}

\vfill \eject

\vskip 1.0cm
\centerline {\bf FIGURE CAPTIONS}
\vskip 0.5cm

\noindent Fig.1. The ratio of the dipole $D$
to quadrupole $Q$ anisotropy of the CMB
as a function of dimensionless matter density $\Omega_{NR}$.
The ratio of the density difference across the horizon $\Delta$
to the density perturbation amplitude $\delta$, which is a small number
for superhorizon perturbations,
serves as a unit of $D/Q$.
(See text for explanations of these parameters). The curves are for
$\Omega_{UR} =10^{-5} $, $10^{-4}$, $10^{-3}$, $10^{-2}$, $10^{-1}$ and
 $1$ from bottom to top.
The cosmological constant is given by
$\lambda=1-\Omega_{NR}-\Omega_{UR}$ in each case  and may be negative.
(a) results for the {\it adiabatic} and (b) {\it isocurvature} perturbations.
(See the text for details).

\vskip 1.0cm

\noindent Fig.2. The evolution of expansion factor $R(t,r)/r$ with time
for $r=0.5$, $1.0$, $1.5$ and $2.0 \times r_0$ (bottom to top).
We use the results obtained for our spherical model with
$\Omega_{NR} = 0.1$ and $\Omega_{UR} = 10^{-4}$.
That illustrates the different expansion rate in different parts of space.

\vfill\eject \end \bye